# AN EXPERIMENTAL APPROACH FOR OPTIMISING MOBILE AGENT MIGRATIONS

Dr . Damianos Gavalas

Department of Cultural Technology and Communication,
University of the Aegean,
Harilaou Trikoupi & Faonos St.,
GR – 81100, Mytilene, Lesvos Island,
Greece

## ABSTRACT

The field of mobile agent (MA) technology has been intensively researched during the past few years, resulting in the phenomenal proliferation of available MA platforms, all sharing several common design characteristics. Research projects have mainly focused on identifying applications where the employment of MAs is preferable compared to centralised or alternative distributed computing models. Very little work has been made on examining how MA platforms design can be optimised so as the network traffic and latency associated with MA transfers are minimised. The work presented in this paper addresses these issues by investigating the effect of several optimisation ideas applied on our MA platform prototype. Furthermore, we discuss the results of a set of timing experiments that offers a better understanding of the agent migration process and recommend new techniques for reducing MA transfers delay.

## 1. INTRODUCTION

Recently, the Mobile Agent (MA) paradigm has emerged within the distributed computing field. The term MA refers to autonomous programs with the ability to move from host to host to resume or restart their execution and act on behalf of users towards the completion of a given task [1].

As a result of the intense research on MA technology, we have witnessed the proliferation of available mobile agent platforms (MAP) over the past few years, currently reaching the impressive number of more than seventy known platforms [2], a lot more than available RPC or CORBA implementations. Among these platforms, some represent commercial initiatives, such as Grasshopper [3] and Aglets [4], whilst others comprise research prototypes expressly oriented to specific application domains, such as network management [5][6][7][8], aiming at optimising flexibility and performance aspects not sufficiently addressed by their general-purpose commercial counterparts.

In general, all these platforms share a set of common design decisions and features. For instance, they all offer an *agency* (i.e. an application that receives, instantiates and provides an execution environment for incoming MA objects), a *migration facility*, a *security component*, a *code repository*, and other software components [9]. Also, with no exception, they have all been developed in Java [10]. Despite the popularity of MAs though, it is today well understood that the employment of MA paradigm in structuring distributed applications is not a panacea, as the consumption of network bandwidth and the delays associated with MA transfers may even result in poorer performance than traditional approaches [11].

To investigate this issue, several research articles have provided quantitative evaluations of the MA paradigm and interesting comparisons against alternative distributed computing paradigms (e.g. CORBA or Java RMI) or client/server- based approaches [5][6][12]. These articles introduced performance models and identified specific application scenarios where the employment of MAs is justified, i.e. results in reduced network traffic and/or latency, and others where alternative approaches are preferable. However, very few works have dealt with the analytical examination of the parameters and design decisions involved in the development of a MAP, aiming at proposing platform design optimisations that could offer migration overhead savings and reduced response time. Furthermore, as far as agent migration delays are concerned, we have not yet come across any research work investigating the contribution of the individual phases involved in every agent transfer on the overall travel latency.

The work presented in this paper is twofold: First, a set of optimisations performed on a MAP research prototype developed by the author are presented; Second, the findings of a number of MA migration response time experiments are discussed, trying to provide a better understanding of agent migrations, thereby identifying ways for reducing their associated latency; to achieve that, the migration process has been broken down to its successive phases, recording the weight of each phase on the overall measured transfer delay.

The paper is organised as follows: Section 2 briefly describes the MAP prototype used to perform agent migration experiments. Platform design optimisations aiming at minimising the MA transfers network overhead and latency are discussed in Sections 3 and 4, respectively. Section 5 presents the results of agent transfer timing experiments. Section 6 reviews related work and Section 7 concludes the paper and draws directions for future work.

## 2. THE MOBILE AGENT PLATFORM

The MAP used to perform agent migration experiments has been entirely developed in Java chosen due to its inherent portability, rich class hierarchy and dynamic class loading capability.

Our framework consists of two major components [9], illustrated in Figure 1:

**(I) Mobile Agents (MAs)**: From our perspective, MAs are Java objects with a unique ID, capable of migrating between hosts where they execute as separate threads and perform their assigned tasks. MAs are supplied with an *itinerary table*, a *data folder* where collected management information is stored and several methods to control interaction with polled devices. Service-oriented MAs, associated with new management tasks, may be created at runtime using the Mobile Agent Generator (MAG) graphical tool. The MAG automatically generates and compiles the code of MAs; MA classes may extend one of the provided super-classes (corresponding to general patterns of tasks).

**(II) Mobile Agent Servers (MAS)**: The interface between visiting MAs and systems is achieved through MAS modules, installed on every device intended to execute MA threads. The MAS logically resides above local legacy system and the Java Virtual Machine (JVM), creating an efficient run-time environment for receiving, instantiating, executing, and dispatching MA objects.

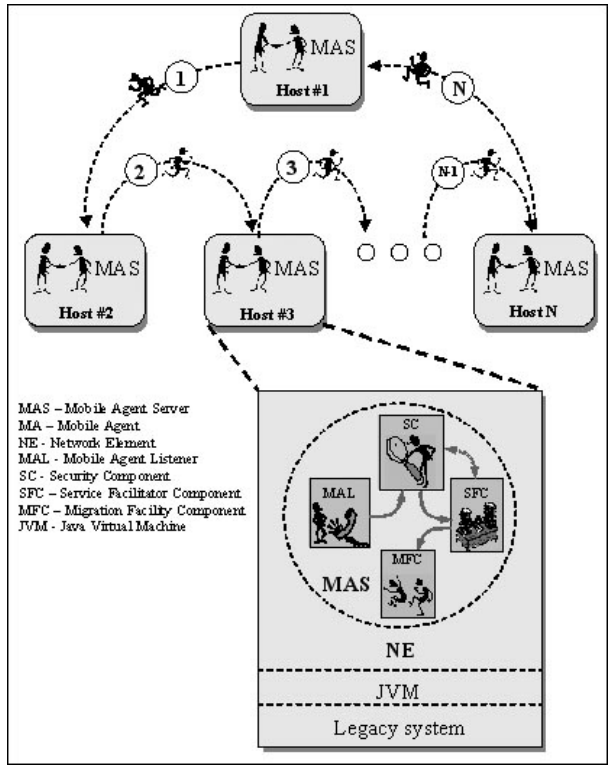

Figure 1. The Mobile Agents-based NM Infrastructure

The MAS also provides requested information to incoming MAs and protects the host system against external attack. The MAS composes four primary components: (a) Mobile Agent Listener, (b) Security Component, (c) Service Facility Component, and (d) Migration Facility Component. Special focus has been given on the design of the Security Component in order to face the security threats represented by executing MAs]. Thus, in addition to the *authorisation* of the MAs requests, the RSA algorithm has been implemented to provide *authentication* of incoming MAs and *encryption* of the obtained sensitive NM data.

## 3. PLATFORM DESIGN CONSIDERATIONS: MINIMISING MOBILE AGENT SIZE

The MA paradigm involves the mobility of a whole computational component from a network device to another, that is, a bundle of *code* and persistent *state* information. For MAPs developed in Java, the *code* part corresponds to the Java bytecode (i.e. the class file), while the *state* part to a set of variables or objects whose values may be altered during the MA travel; these objects may also be used to store the data collected by the agent throughout its lifecycle.

Since the network overhead associated with agent migrations is directly related to the MA size, possible optimisations on MAP design should focus on minimising both the constituting parts of MA components (i.e. code and state). The following two subsections describe design models and programming techniques for minimising the overhead of MA code and state, respectively.

### 3.1. Minimising MA code transfers overhead

Agent code size is typically far larger than its corresponding state size [11], hence, a key objective is to minimise the overhead of MA code transfers. Yet, with the exception of few (e.g. [7], [8]), existing MAPs involve the transfer of both code *and* state on *every* MA migration. The transfer of code though is unnecessary, unless the MA visits a device for a first time, as the Java ClassLoader stores every loaded class on a local code hashtable. That inefficient scheme may result in serious scalability problems both in terms of latency and migration overhead.

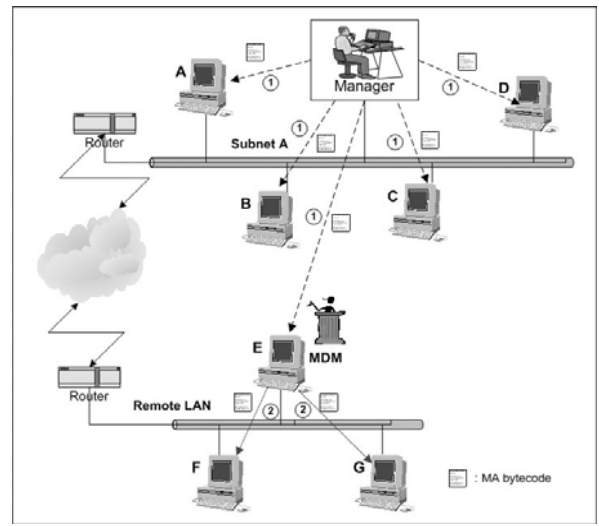

Figure 2. Distribution of MA bytecode through "tree multicasting"

In a previous work [6], we have presented a hierarchical MAP prototype as an extension of the core MAP framework introduced in the preceding section. The hierarchical MAP is particularly suitable for the management of large-scale enterprise networks. The transition to hierarchical agent-based structures is achieved through a mid-level entity, termed the Mobile Distributed Manager (MDM); the MDMs, being mobile agents themselves, operate at an intermediary level between the monitoring device (manager) and the managed hosts and take full control of managing a given network segment. These entities introduce a degree of flexibility as the management system may adapt to changing networking conditions, i.e. MDMs may be dynamically deployed, removed or re-allocated given that specific criteria are met, while also localising the traffic associated with the monitoring of their assigned managed devices.

In monitoring applications as well as in many others (e.g. service management), which can benefit from the employment of the MA paradigm, the set of nodes to be visited by the MA objects (i.e. the agent's itinerary) is known beforehand. Also, itinerant MAs typically visit these nodes more than once. In such applications it is, therefore, clearly unnecessary and inefficient to repeatedly transfer the same agent code to the same devices.

Building on the top of the research prototype described in [6], we have implemented a lightweight mechanism for distributing MA code, implementing the "push" model, defined in [13]. Namely, bytecode is distributed at the MA construction time, as soon as the nodes upon which the MA will perform its tasks (the agent's itinerary) is determined; only the state information is transferred thereafter, resulting in a much lower demand on network resources and faster class loading.

The introduction of the hierarchical model [6] reduces the code distribution cost even further by adopting a "*tree multicasting*" approach. In particular, MA bytecode distribution takes place in two successive phases: In the first phase the MAs bytecode is no longer broadcasted to all managed devices, but instead distributed to the systems local to the manager segment and also to all active MDMs. In the second phase, MDMs will forward the received bytecode to their supervised nodes (see Figure 2). Should a remote domain including *N* hosts is connected to the manager site through a low bandwidth link, the tree multicasting approach will considerably decrease management cost as bytecode is transferred only once (instead of *N* times) through the interconnecting link.

Also, to accelerate the execution of MAs, we have implemented a code pre-loading policy. Namely, every agency maintains a cache of agent bytecodes, pre-loaded at the code distribution phase. Following that, incoming agent objects are instantiated using the cashed bytecode data, with the instantiation being significantly faster (memory access is faster than disk access).

### 3.2. Minimising MA state size

MAs own their mobility characteristic to their ability to carry their persistent state as the latter represents the knowledge/data collected throughout their lifecycle. State information comprises the output of the *serialisation* process [14]. At the destination end, the state is reconstructed through the inverse process of *de-serialisation*. MAs state includes information about the MA class (e.g. the name of the class and the package this class belongs to) and the values of the non-transient and non-static variables/objects declared within the MA class [14] (upon instantiation, transient objects are always assigned their initial values, namely they do not have any 'memory' of their value at the time the MA was serialised). Should this class extends a superclass defining the basic MA functionality, the MA's state will include the values of the non-transient objects declared within the superclass as well.

Since the proposed framework involves *only* the transfer of MA objects state and not their code, it is apparent that the minimisation of the state size is a crucial factor on reducing the overall network overhead. In addition to compressing the MA state using compression algorithms (e.g. the Java *gzip* utility), it is worthwhile investigating alternative techniques for further reducing the volume of transferred data. As a result, a number of experiments/measurements have been conducted and several possible optimisations have been identified; when performing these optimisations, significant cost savings in terms of the state size may be achieved. The list of potential optimisations follows:

**(a) Reduce the number of non-transient objects**: Only the objects whose values are subject to change at each visited host, i.e. the objects that represent the knowledge obtained by an MA during its lifecycle, should be declared as normal (non-transient). The remainder variables should be declared as *transient* and assigned an initial value. Each time the MA is re-instantiated (de-serialised), the transient variable is assigned the same value specified by the programmer, even if that value was changed during the MA's execution on a previously visited host. In other words, when applicable, an object value should be '*hard-coded*' within the MA class. That decreases the degree of flexibility and autonomy given to the MA, as its behaviour/decisions cannot depend on the value of the transient objects, but saves the potentially unnecessary transfer of information related to the value of these objects.

**(b) Use short variable names**: Serialised MA state carries information about the non-transient *variable names* (in addition to their corresponding values). By shortening the length of these names (e.g. using abbreviations instead of long, fully descriptive names) the amount of bytes required to encode this information would be reduced.

| Modification | Size increment for non-compressed state (bytes) | Size increment for compressed state (bytes) |
|---|---|---|
| Use a java.util.Vector structure instead of String[] | 32 | 5 |
| Use java.lang.Integer object instead of int | 19 | 10 |
| Use 20 characters string instead of 3 characters string | 17 | 15 |
| Include an additional 10 characters-long non-transient *String* variable | 47 | 23 |
| Include an additional non-transient *int* variable | 17 | 10 |
| Include an additional non-transient java.util.Vector variable containing a 10 characters-long string | 79 | 28 |
| Include an additional non-transient String[] variable containing a single 10 characters-long string | 49 | 29 |

Table 1. The effect of applying various source code modifications on the ma's state size

**(c) Use short MA class names**: The MA state also embodies information referring to the corresponding MA *class name* as well as the *package* this class belongs to. Similarly to the previous optimisation, by assigning brief names to both the MA class and any packages, a few extra bytes could be saved.

**(d) Use primitive instead of complex data types**: During the serialisation process, referenced objects are processed recursively until all non-transient objects are serialised. When not using primitive data types, but complex data structures for non-transient objects, improved flexibility is offered at the expense of increased state size, as these structures typically contain references to other objects. It is therefore preferable to choose primitive types whenever this is feasible, for instance to use *arrays* instead of `java.util.Vector` structures to store a list of values or to represent integer variables with *int* rather than with powerful and complex `java.lang.Integer` objects.

To illustrate the effect of the optimisations discussed above, we have measured the difference experienced in state size when applying several modifications. These measurements are presented in Table 1.

Significant reductions on the overall state size may be achieved by applying some of the optimisations outlined above. Given that the state size is typically not more than a few hundreds of bytes, it is clear that by summing up the savings obtained by the individual optimisations may lead to trimming a significant portion of state information.

The proposed optimisations are illustrated in Figure 3. The source code of an MA object with 'non-optimised' state is listed in Figure 3a, with its 'optimised' counterpart presented in Figure 3b. It can be noticed that in the latter case, several variables have been declared as transient, simpler data structures are used, whilst variable, class and package names have been shortened. Through applying all these optimisations, the size of the MA state is reduced from 628 bytes (391 when compressed) down to 333 bytes (227 when compressed). That corresponds to a reduction of 47% (41.9% when compressed).

It is clear though that transient objects do not allow the MA to make autonomous decisions based on their value. For instance, the MA corresponding to Figure 3a will execute the `Task()` method only every second hop, while that shown in Figure 3b cannot make such a decision based on the value of transient variable '*hop*', which is assigned the same value *hop = 0* every time the MA arrives at a new host. In addition, non-transient objects make it easier for user applications to interact with the MA and dynamically change its state content at runtime. For instance, in Figure 3a, the user passes to the MA, through its constructor, the name of the originating host to whom the MA will return when completing its itinerary; the name can vary, depending on the host that creates the agent. In contrast, in Figure 3b the name of the originated host is hard-coded within the MA bytecode and not transferred within the MA state. That implies an inflexible design, which is preferable though in the case that MAs are always created by a single device. However, should that device changes, the originating host name will need to be changed accordingly, that is the MA class will have to be modified and *re-compiled*. Concluding, the choice of whether to declare an object as transient or non-transient represents a trade-off between flexibility and migration cost.

```
package MobileAgentPackage;

public class MobileAgentExample extends
                           MobileAgentSuperclass {
    Vector itinerary;
    Vector datafolder = new Vector();
    String originatingHost;
    boolean encryptData;
    boolean doTask = true;
    int hop = 0;

    public MobileAgentExample (Vector it, String host,
                               boolean encrypt) {
        itinerary = it;
        originatingHost = host;
        encryptData = encrypt;
    }

    public void onArriving() {
        hop++;
        doTask = (hop % 2 == 0) ? true : false;
    }

    public void run() {
        System.out.println ("Hop number: " + hop);
        if (doTask)
            Task();
    }

    void Task() {
    }
}
```
(a)

```
package MAPack;

public class MAExample extends MA {
    String[] it;
    String[] data;
    transient String origin = "plato.ct.aegean.gr";
    transient boolean encryptData = true;
    transient boolean doTask = true;
    transient int hop = 0;

    public MAExample (String[] itinerary) {
        it = itinerary;
    }

    public void onArriving() {
        hop++;
    }

    public void run() {
        System.out.println ("Hop number: " + hop);
        if (doTask)
            Task();
    }

    void Task() {
    }
}
```
(b)

Figure 3. Example source code of an MA with (a) 'non-optimised' vs. (b) 'optimised' state size.

# 4. PLATFORM DESIGN OPTIMISATIONS: MINIMISING THE LATENCY OF MOBILE AGENT TRANSFERS

In addition to reducing MA state size, minimising the delay associated with MA transfers is also of major importance, especially when considering time-critical management tasks. The investigation of the migration process itself is presented in Section 5, where the MA transfer is broken down to several successive phases and ways for accelerating agent transfers are identified.

Generally, MA transfers may be realised either over TCP or UDP as transport protocols (although more heavyweight protocols such as Java RMI could also be considered) [12]. Hence, this section aims at investigating ways for optimising the performance of these protocols. The protocol parameters that may be customised through Java network programming are limited to the following:

**(a) Stream buffering**: The hierarchical nature of the I/O classes library allows the programmer to build up streams in hierarchical manner. In particular, by wrapping a `java.io.InputStream` object into a `java.io.BufferedInputStream` instance and the latter into a `java.uil.zip.GZIPInputStream` object, we can take advantage of the `BufferedInputStream` class to enable data buffering and `GZIPInputStream` class to enable data compression when receiving or dispatching MAs. Without buffering, data would be read byte-by-byte, thus degrading performance [15].

**(b) TCP "No Delay" option**: This optimisation refers to MA transfers through TCP. By setting the TCP *NODELAY* option for TCP connections established for transferring MA objects, the transfer is accelerated.

**(c) Further optimisations**: In addition to configuring transport protocol parameters, as suggested above, there are also several improvements that may be performed to speed up the MA migration process. For instance, minimising the MA state size through adopting the optimisations suggested in Section 2 would positively affect the migration latency, as it would considerably reduce the volume of transferred data. A second possible optimisation would be to include IP addresses rather than host names in the MAs itinerary table. That would save the time needed to access the Domain Name Server, in other words, it would obviate the need to translate the names to their corresponding IP addresses prior to MA transfers.

# 5. MOBILE AGENT MIGRATIONS TIMING EXPERIMENTS

In this section, we present the result of timing experiments that aim at providing a better understanding of the response time measured for MA migrations. The objective is to identify the individual phases composing an agent migration and investigate how the overall time is distributed among them as well as the effect of a number of factors such as the transport protocol used on MA transfer delays. In particular, the migration process is first divided into several independent, successive phases in order to assess the weight of their respective delays on the overall latency.

The MA used for this experiment, termed 'ping-pong' MA, is launched at Host_A, visits the agency of Host_B and then returns back to Host_A, *without* performing any kind of task. We have not chosen single-hop itineraries instead of the A→B→A travel, as the former would require precise clock synchronization of source and destination hosts. By using 'ping-pong' agents instead, all time measurements are performed right before the dispatch and right after the arrival of the agent at host A. Therefore, the phases identified are:

1. MA object creation (instantiation);

2. MA object serialisation by the agency of Host_A;

3. MA transfer to Host_B;

4. MA object de-serialisation by Host_B agency;

5. MA object serialisation by Host_B agency;

6. MA transfer back to Host_A;

7. MA object de-serialisation by Host_A agency.

Both TCP and UDP transport protocols are considered for the MA transfers, while the serialised MA state can either be compressed before its transmission or non-compressed. Another investigation issue is the effect of the serialised MA's state size on the overall migration delay. In particular, one of the experiment's objectives is to examine to what extend does the increment of the state size affect the distribution of the individual migration phases delay. The average times corresponding to each migration phase are graphically illustrated in Figure 4 and Figure 5, with the former corresponding to an MA object with relatively moderate state size (476 bytes when compressed, 678 bytes when uncompressed) and the latter to an MA with larger state size (1152 bytes when compressed, 3970 bytes when uncompressed). In both cases, illustrated delays represent the average of 100 individual time measurements. The state size growth is realised by increasing the number of the MA's non-transient variables.

It is noted that all the experiments have been conducted on 10 Mbps Ethernet, using WinNT stations with Pentium III (450 MHz) processors and 128MB of memory.

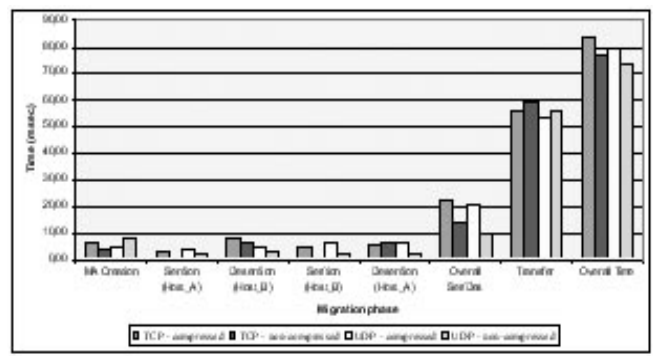

Figure 4. Time measurements depicting the distribution of delays for a 'ping-pong' MA with 'moderate' state size (476 bytes compressed / 678 bytes uncompressed)

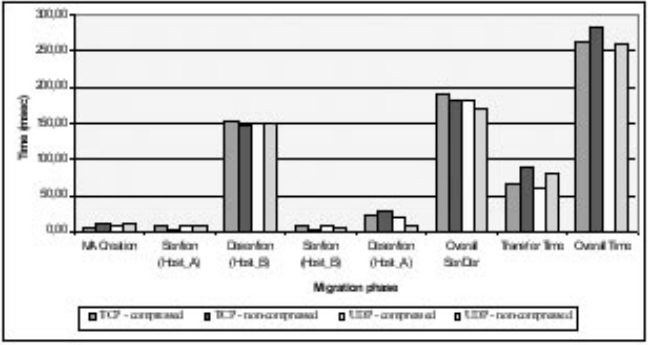

Figure 5. Time measurements depicting the distribution of delays for a 'ping-pong' MA with 'large' state size (1152 bytes compressed / 3970 bytes uncompressed)

It is interesting to observe the way that compression affects the response time as a function of the volume of compressed data. In particular, regardless of the utilised transport protocol (TCP or UDP), MA's state compression always causes reduction of transfer time (the volume of transmitted data is decreased) and increment of the overall serialisation/de-serialisation time (compression/de-compression is regarded as part of the serialisation/de-serialisation process). However, the effect of compression on the overall response time largely depends on the MA's state size. For serialised state of moderate size, compression increases the overall time. Conversely, when the migration of MAs with larger state sizes is considered, compression results in a reduction of the response time. Therefore, compression reduces the latency of the migration process in case that the original size of the serialised state and the compression ratio is such so as to justify the time penalty involved in the compression process. To illustrate, in our first type of MA, compression saves only 678-476 = 202 bytes from being transferred through the network (29.8% of the original data volume), while in the second type the saving becomes 3970 - 1152 = 2818 (71% of the original data volume).

Another interesting aspect of the MA migration time measurements is the alterations that state size causes on the distribution of the overall response time among the individual migration phases. The histograms of Figure 6 confirm that the increase of MA state size results in shifting response time "centre of mass" from transfer to the serialisation/de-serialisation process. For instance, the transfer latency of an MA with moderate state size covers the 66.7% of the overall response time, with the sum of the individual serialisation/de-serialisation times comprising the 26.1%. In contrast, when dealing with the MA with large state size, these percentages become 25.4% and 72.1% respectively.

Interestingly, de-serialisation process is proved more time consuming than its inverse process of serialisation. In addition, a significant portion of the response time is covered by the MA's state de-serialisation, which takes place at the Host_B side (see Figure 6b). This is because we used a customised ClassLoader, which is slower than the default Java ClassLoader used by Host_A application. The serialisation process taking place at Host_B is significantly faster than de-serialisation, since it uses the MA class definition that has been already loaded by the ClassLoader.

In general, the response time difference observed between the two types of MAs is mainly due to the difference in the overall serialisation/de-serialisation time, as shown in Figure 7. Certainly, the transfer time also increases in the case of the MA with larger state size; however, this is not the decisive factor.

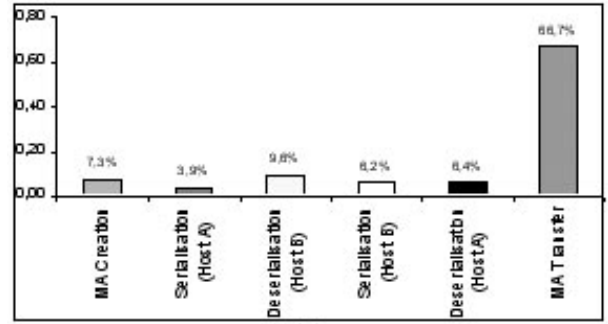

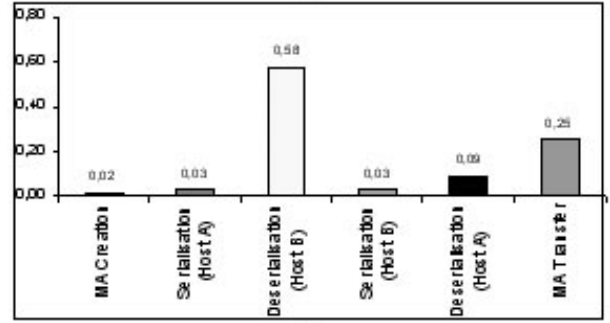

Figure 6. Distribution (percentage of the overall response time) of the delays incurred within the individual migration phases for MAs with (a) moderate or (b) fairly large state size (the serialised state is compressed)

A possible optimisation would be to save the time needed to create (instantiate) the MA object, as suggested in Section 4. That could be realised through requesting Host_A to create the MA before the actual trip starts. Such an optimisation would result in performance gain, especially in the case of MAs with small state size, when MA creation represents a considerable proportion of the overall time (7.3%).

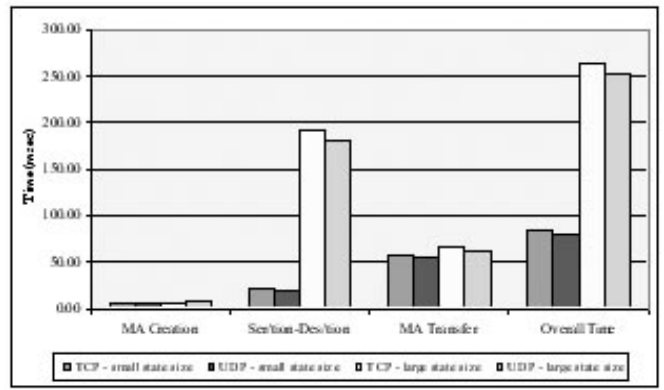

Figure 7. Comparison of the individual migration phases delays for MAs with moderate vs. large state size (the serialised state is compressed prior to its transmission)

## 6. RELATED WORK

Most of the existing MAPs do not implement any special techniques to achieve high-performance and lightweight agent migrations.

The main exception is James platform [8] which implements a flexible code distribution mechanism termed 'code prefetching'. Within this approach, all agencies included into an agent's itinerary are informed about this anticipated visit. Using this information, the agencies perform the code loading in advance, while the agent is still executing in a previous place of its itinerary.

Puliafito et al. [7] proposed an alternative class loading mechanism where agencies load incoming MAs code on demand, when the MA object visits the hosting device for a first time. From our point of view though, our approach represents a more time efficient solution as the MA code is distributed to the network devices immediately after the agent's itinerary is determined. As a result, the MA starts executing its tasks right after being received by the visited agency as its code is pre-loaded and maintained in the local system's cache. In any other case, the time required to contact a code repository and download the required bytecodes may lead to unacceptable delays, especially when considering time critical tasks.

However, we have not yet come across any research work proposing ways to minimize itinerant MAs persistent state size, although utilizing MAs with large state may severely affect the performance and scalability of MA-based applications. In terms of timing experiments, several articles included measurements of migration delays for specific agent platforms (e.g. [5], [12]), while others presented interesting comparative performance, robustness and scalability tests for the most popular commercial MAPs [16]. Yet, to the best of our knowledge, experimental results investigating the weight of the individual phases of an agent migration on the overall latency have not yet been reported.

It should be stressed that a comparison among our MAP prototype and other available MAPs does not fall within the scope of this article for two reasons: (a) our main objective is not to suggest the most lightweight, high-performance or stable amongst the available platforms, but instead to provide a better understanding of agents migration process and also to identify ways for minimizing the associated latency and incurred traffic; (b) such a comparison would presuppose that the experimentally tested MAPs would be publicly available and -most importantly- open source projects (otherwise, timing experiments would be restricted on measuring the overall migration delay, since the "separation" of individual MA migration phases would not be possible); open-source MAPs though, are very few (e.g. [4][17]). Besides, the design considerations detailed in the article as well as the conclusions extracted from the experimental results are generic enough and do not only apply to the author's research prototype.

## 7. CONCLUSIONS & FUTURE WORK

The phenomenal intensity of research on MA technology is reflected on several industrial and academic initiatives that led to the development of numerous MAPs. Research prototypes, analytical performance models and quantitative evaluations are already out there; applications scenarios that may benefit from the employment of this technology have also been identified. However, not much has been done on examining how MAPs design could be optimised so as to minimise the network traffic and delays associated with MA transfers. The work presented in this paper addressed these issues by proposing and investigating the effect of several optimisations applied on our MAP prototype.

Furthermore, we performed a set of agent migration timing experiments; the findings of these experiments revealed the impact of individual migration process phases, thereby identifying ways for reducing the associated latency.

As a future work, in order to minimise the usage of distributed system resources we plan to implement an efficient code replacement policy; namely, to remove loaded MA classes from systems cache or even from the local disk space according to a 'least recently used' algorithm when the usage of local resources exceeds a specified threshold.

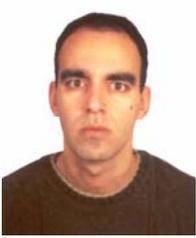

Dr. Damianos Gavalas received his BSc degree in Computer Science from University of Athens, Greece, in 1995 and his MSc and PhD degree in Electronic Engineering from University of Essex, U.K., in 1997 and 2001, respectively. He has been employed as research engineer in telecommunication and IT industries.

In July 2004 he was appointed as a Lecturer in the Department of Cultural Technology and Communication, University of the Aegean, Greece. His research interests include distributed computing, mobile code, network and systems management, e-commerce, mobile ad-hoc networks multicasting and routing. He is a member of the IEEE Computer and Communication societies.